# Snap-through instability of graphene on substrates

Teng Li[*], Zhao Zhang

Department of Mechanical Engineering, University of Maryland, College Park, MD 20742

Maryland NanoCenter, University of Maryland, College Park, MD 20742

## Abstract

We determine the graphene morphology regulated by substrates with herringbone and checkerboard surface corrugations. As the graphene/substrate interfacial bonding energy and the substrate surface roughness vary, the graphene morphology snaps between two distinct states: 1) closely conforming to the substrate and 2) remaining nearly flat on the substrate. Such a snap-through instability of graphene can potentially lead to desirable electronic properties to enable graphene-based devices.

[*] Corresponding author. Email: LiT@umd.edu

Graphene is a monolayer of carbon atoms densely packed in a honeycomb crystal lattice. It exhibits extraordinary electrical and mechanical properties[1], and has inspired an array of tantalizing potential applications (e.g., transparent flexible displays[2] and biochemical sensor arrays[3]). Graphene is intrinsically non-flat and tends to be randomly corrugated.[4] The random graphene morphology can lead to unstable performance of graphene devices as the corrugating physics of graphene is closely tied to its electronic properties.[5] Future success of graphene-based applications hinges upon precise control of the graphene morphology over large areas, a significant challenge largely unexplored so far. Recent studies show that, however, the morphology of graphene can be regulated by the surface of an underlying substrate.[6] In this paper, we quantitatively determine the regulated graphene morphology on substrates with various surface patterns, using energy minimization. The results reveal the snap-through instability of graphene on substrates, a promising mechanism to enable functional components for graphene devices.

Recent experiments show that monolayer and few-layer graphene can partially follow the rough surface of the underlying substrates.[6] The resulting graphene morphology is regulated, rather than the intrinsic random corrugations in freestanding graphene. The substrate-regulated graphene morphology results from the interplay between the interfacial adhesion and the strain energy of the graphene/substrate system, which can be explained as follows.

When graphene is fabricated on a substrate surface via mechanical exfoliation[7] or transfer printing[8], the graphene/substrate interfacial adhesion is usually weak (e.g., van der Waals interaction). As the graphene corrugates to follow the substrate surface, the graphene/substrate interaction energy decreases due to the nature of van der Waals interaction; on the other hand, the strain energy in the system increases due to the intrinsic bending rigidity of graphene. At the equilibrium graphene morphology on the substrate, the sum of the interaction energy and the system strain energy reaches its minimum.

The above energetic consideration can be used to quantitatively determine the regulated graphene morphology on a rough substrate surface. Furthermore, with a systematic understanding of the governing mechanisms of substrate-regulated graphene morphology, we envision a promising strategy to precisely pattern graphene into desired morphology on engineered substrate surfaces. In this paper, we illustrate this strategy by determining the regulated graphene morphology on two types of engineered substrate surfaces: herringbone corrugation and checkerboard corrugation (Fig. 1). These substrate surface features can be fabricated via approaches combining lithography[9] and strain engineering[10].

The graphene/substrate interaction energy, denoted by $E_{\text{int}}$, can be estimated by summing up all van der Waals forces between the graphene carbon atoms and the substrate atoms. The van der Waals force between a graphene-substrate atomic pair of distance $r$ can be characterized by a Lennard-Jones pair potential, $V_{LJ}(r) = 4\varepsilon(\sigma^{12}/r^{12} - \sigma^{6}/r^{6})$, where $\sqrt[6]{2}\sigma$ is the equilibrium distance of the atomic pair and $\varepsilon$ is the bonding energy at the equilibrium distance. We have developed a Monte Carlo numerical scheme to compute the graphene/substrate interaction energy.[11]

The strain energy in the graphene/substrate system results from the corrugating deformation of the graphene and the interaction-induced deformation of the substrate. When an ultrathin monolayer graphene partially conforms to a rigid substrate (e.g., SiO$_2$), the substrate deformation due to the weak graphene/substrate interaction is expected to be negligible. Therefore in this paper we neglect the substrate strain energy. Also, when the graphene spontaneously follows the substrate surface under weak interaction and is not subject to any mechanical constraints (e.g., pinning[12]), the in-plane stretching of the graphene is also expected to be negligible. Therefore, in this paper we only consider the graphene strain energy due to out-of-plane bending, denoted by $E_g$. Effect of the above assumptions on results is to be further elaborated in the latter part. Denoting the out-of-plane displacement of the graphene by $w(x, y)$, the graphene strain energy per unit area over its area $S$ can be given by

$$E_g = (1/S)\int_S D[(w_{,xx} + w_{,yy})^2/2 - (1-\nu)(w_{,xx}w_{,yy} - w_{,xy}^2)]dS, \quad (1)$$

where the subscripts of $w$ denote differentiation, and $D$ and $\nu$ are the bending rigidity and the Poisson's ratio of graphene, respectively.

The out-of-plane herringbone corrugation of the substrate surface (Fig. 1a) and the out-of-plane corrugation of the graphene regulated by such a substrate surface are described by

$$\begin{aligned} w_s &= A_s \cos\left((2\pi/\lambda_x)(x + A_y \cos(2\pi x/\lambda_y))\right) - h \\ w_g &= A_g \cos\left((2\pi/\lambda_x)(x + A_y \cos(2\pi x/\lambda_y))\right) \end{aligned}, \quad (2)$$

respectively, where $A_s$ and $A_g$ are the amplitudes of the substrate surface corrugation and the graphene corrugation, respectively; for both the graphene and the substrate, $\lambda_x$ is the wavelength of the out-of-plane corrugation, $\lambda_y$ and $A_y$ are the wavelength and the amplitude of in-plane jogs, respectively; and $h$ is the distance between the middle planes of the graphene and the substrate surface.

For a given substrate surface corrugation (i.e., $A_s$, $A_y$, $\lambda_x$ and $\lambda_y$), $E_g$ increases monotonically as $A_g$ increases. On the other hand, $E_{int}$ minimizes at finite values of $A_g$ and $h$, due to the nature of van der Waals interaction. As a result, there exists a minimum of $(E_g + E_{int})$ where $A_g$ and $h$ reach their equilibrium values. The energy minimization is carried out by running a customized code on a high performance computation cluster. In all computations, $D = 1.41\,eV$, $\sigma = 0.353\,nm$ and $A_s = 0.5\,nm$, which are representative of a graphene-on-SiO$_2$ structure.[13] Various values of $\varepsilon$, $\lambda_y$, $\lambda_x$ and $A_y$ are used to study the effects of interfacial bonding energy and substrate surface pattern on the regulated graphene corrugation.

Figures 2a plots the normalized amplitude of the regulated graphene corrugation, $A_g / A_s$, as a function of $D/\varepsilon$ for various $\lambda_x$. Here $\lambda_y = 2\lambda_x$ and $A_y = \lambda_x/4$. Thus various $\lambda_x$ define a family of substrate surfaces with self-similar in-plane herringbone patterns and the same out-of-plane amplitude (i.e., $A_s$). For a given substrate surface pattern, if the interfacial bonding energy is strong (i.e., small $D/\varepsilon$), $A_g$ tends to $A_s$. In other words, the graphene closely follows the substrate surface (Fig. 2b). In contrast, if the interfacial bonding is weak (i.e., large $D/\varepsilon$), $A_g$ approaches zero. That is, the graphene is nearly flat and does not conform to the substrate surface (Fig. 2c). Interestingly, there exists a threshold value of $D/\varepsilon$, below and above which a sharp transition occurs between the above two distinct states of the graphene morphology. We call such a sharp transition *the snap-through instability of the graphene*. The threshold value of $D/\varepsilon$ increases as $\lambda_x$ increases. For a given interfacial bonding energy, $A_g$ increases as $\lambda_x$ increases. That is, graphene tends to conform more to a substrate surface with smaller out-of-plane waviness.

Further simulations (not shown in Fig. 2) show that, for a given interfacial bonding energy, if $\lambda_x$ and $A_y$ are fixed, $A_g$ increases as $\lambda_y$ increases; and if $\lambda_x$ and $\lambda_y$ are fixed, $A_g$ increases as $A_y$ decreases. That is, graphene tends to conform more to a substrate surface with smaller in-plane waviness. Moreover, the snap-through instability of the graphene similar to that illustrated in Fig. 2 is also evident in these simulations.

The snap-through instability of the graphene on a substrate surface can be explained as follows. Figure 3 plots the normalized total system energy as a function of $A_g/A_s$ for various $D/\varepsilon$. Here $\lambda_x = 9nm$, $\lambda_y = 2\lambda_x$ and $A_y = \lambda_y/4$. If the interfacial bonding energy is weaker ($D/\varepsilon = 575$) than a threshold value, the total energy profile reaches its minimum at a small graphene corrugation amplitude $A_g/A_s = 0.14$. If the interfacial bonding energy ($D/\varepsilon = 750$) is stronger than the threshold value, the total energy profile reaches its minimum at a large graphene corrugation amplitude $A_g/A_s = 0.93$. At the threshold value of $D/\varepsilon = 650$, the total energy profile assumes a double-well shape, whose two minima ($A_g/A_s = 0.20$ and $0.91$) correspond to the two distinct states of the graphene morphology on the substrate surface.

In the case of graphene regulated by a substrate surface with checkerboard pattern (Fig. 1b), the substrate surface corrugation and the regulated graphene corrugation are described by

$$w_s = A_s \cos(2\pi x/\lambda)\cos(2\pi y/\lambda) - h$$
$$w_g = A_g \cos(2\pi x/\lambda)\cos(2\pi y/\lambda) \qquad (3)$$

respectively, where $\lambda$ is the wavelength of the out-of-plane corrugation for both the graphene and the substrate surface.

Figure 4 plots $A_g/A_s$ on the checkerboard substrate surface as a function of $D/\varepsilon$ for various $\lambda$. For a given substrate surface roughness, $A_g/A_s$ decreases as $D/\varepsilon$ increases. For a given interfacial bonding energy, $A_g/A_s$ increases as $\lambda$ increases. On a substrate surface with checkerboard corrugation, graphene exhibits the snap-through instability as well, which also results from the double-well shape of the system energy profile at the threshold value of $D/\varepsilon$, similar to that shown in Fig. 4. The threshold value of $D/\varepsilon$ at the graphene snap-through instability increases as $\lambda$ increases.

In this paper we focus on graphene morphology spontaneously regulated by substrate surfaces via weak interaction. When a graphene/substrate structure is subject to external loading, the graphene strain energy due to stretching and the substrate strain energy may also need to be considered. In this sense, the present model overestimates the graphene corrugation amplitude. Also the graphene/substrate interaction can be enhanced by the possible chemical bondings or pinnings at the interface.[12,14] In this sense, the present model underestimates the graphene corrugation amplitude.

In summary, we investigate the graphene morphology regulated by substrates with herringbone and checkerboard surface corrugations. Depending on interfacial bonding energy and substrate surface roughness, the graphene morphology exhibits a sharp transition between two distinct states: 1) closely conforming to the substrate surface and 2) remaining nearly flat on the substrate surface. The quantitative results suggest a promising strategy to control the graphene morphology through substrate regulation. While it is difficult to directly manipulate freestanding graphene[15], it is feasible to pattern the substrate surface via lithography[9] and strain engineering[10]. The regulated graphene morphology on such engineered substrate surfaces may lead to ways to control the graphene electronic properties, introducing desirable properties such as band-gap, or p/n junction behavior. For example, the graphene snap-through instability on substrates can possibly enable the design of graphene nano-switches. We then call for experimental demonstration.

This work is supported by the Minta-Martin Foundation. Z.Z. also thanks the support of the A. J. Clark Fellowship.

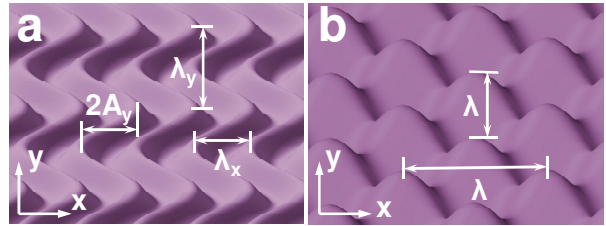

**Fig. 1.** (Color online) Schematics of substrate surface corrugations: (a) herringbone and (b) checkerboard.

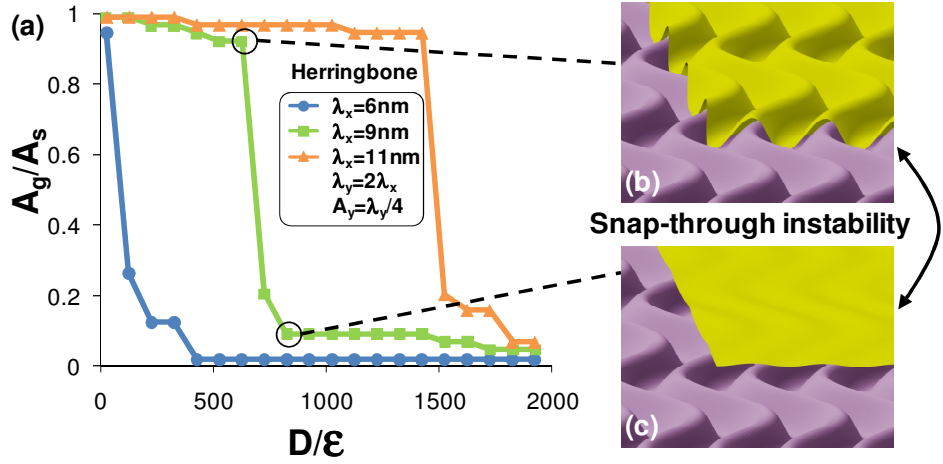

**Fig. 2.** (Color online) (a) $A_g / A_s$ on substrates with herringbone surface corrugation as a function of $D/\varepsilon$ for various $\lambda_x$. At a threshold value of $D/\varepsilon$, the graphene morphology snaps between two distinct states: (b) closely conforming to the substrate surface and (c) remaining nearly flat on the substrate surface.

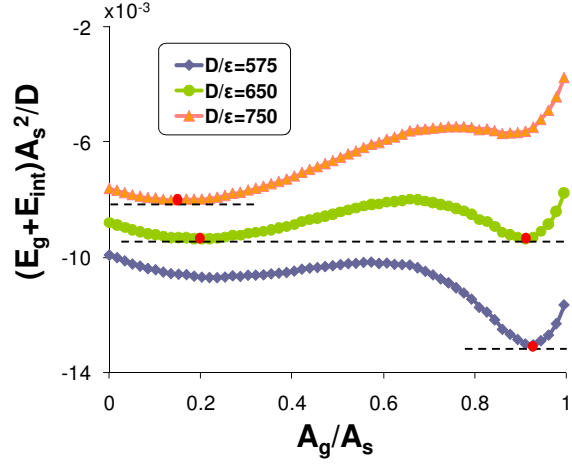

**Fig. 3.** (Color online) The normalized total system energy as a function of $A_g/A_s$ for various $D/\varepsilon$. At a threshold value of $D/\varepsilon$, the total system energy minimizes at two points, corresponding to the two distinct states of graphene morphology.

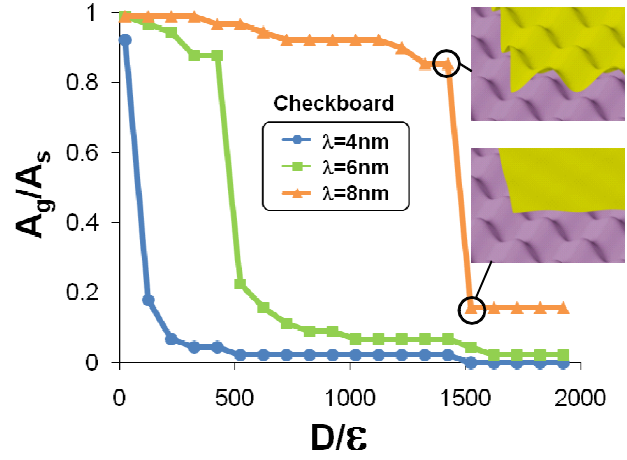

**Fig. 4.** (Color online) $A_g / A_s$ on substrates with checkerboard surface corrugation as a function of $D/\varepsilon$ for various $\lambda$. The insets illustrate the two distinct states of graphene morphology at the snap-through instability.